\documentclass[preprint2]{aastex}
\usepackage{lscape} 
\usepackage{amsmath, amsthm,amssymb}
\usepackage{multirow}
\usepackage{url}

\title{Kuiper Belt Occultation Predictions}

\author{Wesley C. Fraser {$^1$}, Stephen Gwyn {$^1$}, Chad Trujillo {$^2$}, Andrew W. Stephens{$^2$}, JJ Kavelaars {$^1$}, Michael E. Brown{$^3$}, Federica B. Bianco{$^4$}, Richard P. Boyle{$^5$}, Melissa J. Brucker{$^6$}, Nathan Hetherington{$^7$}, Michael Joner{$^{8}$}, William C. Keel{$^{9}$}, Phil P. Langill{$^{10}$}, Tim Lister{$^{11}$}, Russet J. McMillan{$^{12}$}, Leslie Young{$^{13}$}}

\altaffiltext{1}{Herzberg Institute of Astrophysics, 5071 W. Saanich Rd. Victoria, BCV9E 2E7}
\altaffiltext{2}{Gemini Observatory, Northern Operations Center, 670 N A'ohoku Place, Hilo, HI 96720}
\altaffiltext{3}{California Institute of Technology, Division of Geological and Planetary Sciences, 1200 E. California Blvd., Pasadena, CA. 91101 USA}
\altaffiltext{4}{Center for Cosmology and Particle Physics, Department of Physics, New York University, 4 Washington Place, New York, NY 10003, USA}
\altaffiltext{5}{Vatican Observatory Research Group, Steward Observatory, University of Arizona, Tucson, AZ 85721, USA}
\altaffiltext{6}{University of Nebraska - Lincoln, Dept. of Physics \& Astronomy, Jorgensen Hall, Rm 208, 855 N 16th Street, Lincoln, NE 68588-0299}
\altaffiltext{7}{Department of Astronomy and Astrophysics, University of Toronto, Toronto, ON M5S 3H4, Canada}
\altaffiltext{8}{Department of Physics and Astronomy, N283 ESC, Brigham Young University, Provo, UT 84602-4360, USA}
\altaffiltext{9}{Box 870324, University of Alabama, Tuscaloosa, AL 35487-0324, USA}
\altaffiltext{10}{Rothney Astrophysical Observatory, University of Calgary, Calgary AB Canada, T2N 1N4}
\altaffiltext{11}{'Las Cumbres Observatory Global Telescope Network (LCOGT), 6740 Cortona Drive Suite 102 Goleta, CA 93117, USA}
\altaffiltext{12}{Apache Point Observatory, PO Box 59, Sunspot, NM 88349}
\altaffiltext{13}{Southwest Research Institute, 1050 Walnut St., Suite 300, Boulder, CO 80302}

\email{wesley.fraser@nrc.ca}

\date{} 

\slugcomment{Submitted to ...}
\received{É}
\revised{É}

\begin{abstract}
Here we present observations of 7 large Kuiper Belt Objects. From these observations, we extract a point source catalog with $\sim0.01"$ precision, and astrometry of our target Kuiper Belt Objects with $0.04-0.08"$ precision within that catalog. We have developed a new technique to predict the future occurrence of stellar occultations by Kuiper Belt Objects. The technique makes use of a maximum likelihood approach which determines the best-fit adjustment to cataloged orbital elements of an object. Using simulations of a theoretical object, we discuss the merits and weaknesses of this technique compared to the commonly adopted ephemeris offset approach. We demonstrate that both methods suffer from separate weaknesses, and thus, together provide a fair assessment of the true uncertainty in a particular prediction. We present occultation predictions made by both methods for the 7 tracked objects, with dates as late as 2015. Finally, we discuss observations of three separate close passages of Quaoar to field stars, which reveal the accuracy of the element adjustment approach, and which also demonstrate the necessity of considering the uncertainty in stellar position when assessing potential occultations. 
\end{abstract}

\begin{document}

\maketitle

\section{Introduction \label{sec:Intro}}
Detection of a point-source occultation by a planetesimal can provide a wealth of detail about that body. For example, observations of stellar occultations by Kuiper Belt Objects (KBOs) have been used to measure sizes \citep{Elliot2010,Sicardy2011}, shapes \citep{Braga-Ribas2011}, and atmospheric extents \citep{Hubbard1988,Elliot1989} of the occulting KBOs. As the angular extents of most KBOs are $\lesssim0.03"$, even the largest few are only partially resolved in the highest resolution telescope imaging possible \citep{Brown2004b}. As a result, the information gleaned from an occultation about the occulting body is currently impossible to get by any other means. 

The small angular extents of KBOs make it extremely challenging to predict when they will cause an occultation. The best stellar catalogs currently available typically have astrometric accuracies up to an order of magnitude larger than the apparent diameters of the largest KBOs, e.g., $\sim0.25"$ for USNO B1 \citep{Monet1998} and $\sim0.07"$ for the 2MASS catalog \citep{2mass}. For reference, 0.1" uncertainty in stellar position projected to typical KBO distances of 40~AU results in approximately 3000~km uncertainty in the predicted shadow path. Thus, the use of standard astrometric catalogs can result in path uncertainty larger than the Earth.  As done by \citet{Stone1999}, the generation of a custom point-source catalog is the first step required in producing occultation predictions with any certainty. 

An additional source of uncertainty, and equally important as stellar position, is the ephemeris of the KBO in question. As discussed by \citet{Stone1999} for the object (5145) Pholus, ephemeris uncertainty can be as large as 0.5" and can nullify any efforts made in producing accurate point-source catalogs. \citet{Assafin2012} present a method by which the nominal ephemeris of an object - evaluated from the astDys orbital elements\footnote{\url{http://hamilton.dm.unipi.it/astdys/}} -  is corrected by a constant vector value which is measured with respect to their point source catalog. This offset is then used to approximately correct the ephemeris to that catalog. 

The method presented by \citet{Assafin2012} has been successful in predicting a few detected occultations \citep[see, for example,][]{Braga-Ribas2011}. This method however, still suffers from uncertainty in the object's ephemeris. As is shown in Figure~\ref{fig:ephem_comparison}, the use of a nominal ephemeris can result in oscillations between the predicted and actual position of the body as large or larger than a few tens of milli-arcseconds,  the apparent diameters of the largest KBOs.

Here we present a method of occultation prediction similar in fashion to that used by \citet{Assafin2012}. Rather than adopting a constant ephemeris offset, our method uses high precision astrometry to correct the nominal orbital elements themselves. Ephemeris uncertainty (and orbital element uncertainty) are natural end-products of our method, unlike other methods which do not produce a formal ephemeris uncertainty. This results in independent and complementary predictions to those of the constant offset method and a means of assessing the true uncertainty in a given prediction. With this method we present the results of a pilot study to test the feasibility of our approach. We present a  list of candidate stellar occultations for 7 KBOs, spanning as late as 2015.

In Section 2 we present the general method of occultation prediction. We describe our observations, and our method of point-source catalog production. We also describe the method of ephemeris generation, and the net uncertainties resultant from our method. In Section 3 we present our occultation predictions and present some observations which confirm the validity of the method. We finish with concluding remarks in Section 4.

\section{Observations and Predictions\label{sec:observations}}
In this section we present our observations and predictions method. This program was a pilot study of 7 large KBOs: (50000) Quaoar, (84522) 2002 TC302, (90377) Sedna, (136199) Eris, (136472) Makemake, (202421) 2005 UQ513, and (225088) 2007 OR10. These objects are all well tracked KBOs and have ephemerides determined from reported observations spanning many years. The ephemerides of these objects are typical of the most accurate KBO ephemerides available. As a result, these targets should present a simple test case in which only small corrections to their ephemerides will be necessary for occultation prediction purposes.

The first step in predicting stellar occultations is to generate an astrometrically accurate point-source catalog. For this purpose, observations were taken with MegaPrime on the 3.6~m Canada-France-Hawaii Telescope (CFHT). MegaPrime is a 36 CCD optical imager that provides a  fully subsampled (0.1875 "/pixel)  $1\times1^o$ field of view. Observations were taken utilizing 45~s exposures in SDSS r'-filter and spanned a region large enough to ensure that the target of interest was within the Right ascension (RA) and Declination (Dec) range throughout 2011 and 2012. Details of the MegaPrime sky coverage can be found in Table~\ref{tab:CFHT_coverage}.

As a result of the CFHT Legacy Survey\footnote{\url{http://www.cfht.hawaii.edu/Science/CFHTLS/}} \citep{gwyn2008} the spatial distortions of the MegaPrime field of view are very well understood, and can be well described by a global distortion map with only first and second order terms in radius, as well as individual linear distortion maps for each of the chips in the mosaic. This allows us to determine accurate astrometric distortion maps of the images to be produced. It is from these detrended images that our master point-source catalog (MPSC) is assembled. We describe how this is done in Section~\ref{sec:pscp}. 

The MegaPrime images often contain the KBO of interest, and as such also provide some astrometry of the source from which the ephemeris can be corrected. Further tracking observations have been acquired with Gemini Multi-Object Spectrograph \citep[GMOS,][]{Hook2004} on the 8~m Gemini-North telescope. Details of the observations presented here are shown in Table~\ref{tab:observations}. These observations are used to correct the nominal orbital elements of each object to produce an extremely accurate ephemeris in the astrometric system defined by the MPSC. We describe how the element corrections are determined in Section~\ref{sec:elem_corr}.

\subsection{Point-Source Catalog Production \label{sec:pscp}}
Software from the MegaPipe data pipeline \citep{gwyn2008} was used to
produce the point-source catalog and to astrometrically calibrate each
of the MegaPrime images. Starting with images already preprocessed by the Elixir pipeline \citep{Magnier2004}, for each image we produced a source catalog
with positions in pixel coordinates using SExtractor
\citep{hihi}.  SExtractor's parameters were set such that only
fairly bright sources were detected; the detection criteria are set to
flux levels 5 sigma above the sky noise in at least 5 contiguous
pixels. The catalogs were further cleaned of cosmic rays and extended
sources, leaving only point sources. Source centroids were found using SExtractor's  simple centroid method. More complicated methods such as
Gaussian or PSF fitting were found to provide no noticeable benefit.

Each of the pixel coordinate catalogs were matched to an external
astrometric catalog, either the SDSS or 2MASS \citep{2mass}. The SDSS DR9
\citep{sdssdr9} provides a superior source density, and very small
astrometric errors (0.07-0.09"). Where it was not available, 2MASS was chosen as
the reference catalog (astrometric error 0.08-0.09"). We did not use UCAC \citep{ucac2} because it is
fairly shallow; most of the sources are saturated in MegaPrime images,
leaving too few sources for accurate astrometric calibration.  The
USNO catalogs were also considered: they go deeper and have a higher source
density. The astrometric errors (0.4-0.6") on each source are larger than with the 2MASS catalog.
Empirically, it was found that using the 2MASS catalog gave the
smallest astrometric residuals.

Initially, the pixel coordinate catalogs were matched to the external
astrometric catalog on a chip-by-chip basis. The initial Elixir
\citep{elixir} pre-processing done by CFHT provides an initial
astrometric calibration which is typically accurate to better than 1
arcsecond. The matching is therefore relatively simple. Sophisticated
techniques (such as the quad-matching method used by
\citet{astrometry.net}) are not required. During the initial matching process, any catalog source that was more than 1" away from the nearest observe source was ignored. Once the catalogs were
matched, the transformation was computed. For the initial match, we
used a second-order polynomial in x and y. This transformation was
used to refine the matching of sources in the images to sources in the
external catalog, and the transformation was re-computed. The second
transformation was computed slightly differently. The MegaPrime
distortion map can be adequately described by a polynomial with second
and fourth order terms in measured radius, $r$, measured from the center of the
mosaic. This is given by

\begin{equation}
R=r(1+a_{\textrm{1}}r^2+a_{\textrm{2}}r^4),
\end{equation}
\noindent
where $R$ is the true radius, and $a_{\textrm{1}}$ and $a_{\textrm{2}}$ are distortion coefficients.
The coefficients of this polynomial were determined for all 36
chips simultaneously. In addition, a linear distortion map was
computed for each chip. The combination of a global, non-linear
transformation and 36 local, linear transformations sufficiently
describes the distortion, such that additional complexity in the map does not detectably improve the solution. The global transformation takes care of most of
the distortion caused by the MegaPrime optics, while the linear
transformation takes care of any non-coplanarity of the detector array
as well as the effects of differential refraction. We avoid the usually adopted method of determining the full distortion map
which uses a third-order polynomial
transformation for each chip. This utilizes up to 20 parameters per chip,
or a total of 720 parameters for all 36 chips. This number of
parameters is uncomfortably close to the number of sources available
for astrometric calibration. In most MegaPrime fields of view
one typically finds 2000 suitable sources, but the number can be as
low as 1000. 

Once transformations had been determined, they were used to convert
the pixel coordinates in the individual source catalogs to RA and Dec.
The catalogs from each image were merged to produce a master catalog.
For each source from each catalog, all the catalogs are checked for
matching sources. A match occurs if a source in one catalog lies within 2'' of a source in another catalog. The matching sources must also
have measured magnitudes within 1 magnitude of each other. Once all
the matching sources had been found, their measured positions from the
different catalogs were averaged. To avoid confusion in the matching, if two
sources in the same catalog lie within 4'' of each other, both
sources are discarded.

Next, the astrometric calibration of each input image was repeated
using the merged master catalog as the astrometric reference. Using
these new calibrations, the source catalogs for each image are
converted from pixel coordinates to RA and Dec. These catalogs are
merged as described previously to produce a new merged master catalog.
This second catalog is used to calibrate the images a third time.
We refer to this step as ``merge-by-catalog''.

The images were photometrically calibrated by one of three methods:

\noindent
1) If the images overlapped
the SDSS, it was used for calibration. To account for slight differences in photometric systems do to detector and filter differences, the SDSS photometry was
transformed to the MegaPrime system as follows:
\begin{equation}
r_{\textrm{Mega}}= r_{\textrm{SDSS}} - 0.024 (g_{\textrm{SDSS}} - r_{\textrm{SDSS}}).
\end{equation}
This results in an absolute photometric calibration accurate to about 0.01 mag.

\noindent
2) If the image lay outside the SDSS, the Elixir photometric calibration
was used for data taken on photometric nights. The absolute photometric
calibration in this case is slightly worse, typically 0.03 mags.

\noindent
3) Images taken on non-photometric nights were calibrated using parts
of the image which overlapped with images photometrically calibrated
with one of the previous methods.

The photometrically calibrated catalogs were merged using a simliar
method to the astrometric catalog merging described above to produce a
merged master photometric catalog . The final photometric calibration
was done using this catalog. Merging the photometric catalogs
and re-calibrating in this way does not significantly improve the
external photometric calibration, but ensures that the internal
image-to-image zero-point calibration is typically better than 0.005
magnitudes.

The astrometrically and photometrically calibrated images were then
combined using SWarp \citep{swarp}. SWarp is a program that resamples and stacks multiple images onto a projection defined by the image World Coordinate Systems. The background was computed on a
128-pixel grid using a median filter and removed. The astrometric
distortion was removed and the photometric scaling was applied. The
images were resampled to a common grid using a Lanczos 3-pixel
kernel. The scaled, resampled pixels are combined using a median.  We
refer to this step as ``merge-by-pixel''.  SExtractor was then run on the
resulting image to produce the final catalog of point sources which we refer to as the Master Point Source Catalog (MPSC).
Finally, the individual images were re-calibrated using the MPSC as an astrometric reference.

After each iteration of the astrometric calibrations, the image-to-image astrometric residuals were checked.
After the initial match to the external catalog, the astrometric
residuals range from 0.08 to 0.1 arcseconds RMS, typically slightly
higher if the 2MASS catalog was used as an external reference,
slightly lower if the SDSS was used. After using the second master
catalog, the one generated using the ``merge by catalog'' method,
the astrometric residuals are $\sim$0.06 arcseconds. The residuals get
marginally lower if the merge-recalibrate-merge cycle is repeated.
After using the final ``merge-by-pixel'' master catalog the
astrometric residuals are typically 0.03 to 0.04 arcseconds between
two individual images. The residuals between the individual images
and the corresponding master catalog are typically 60-70\% smaller
than residuals between two individual images.

\begin{figure}[h]
   \centering
   \plotone{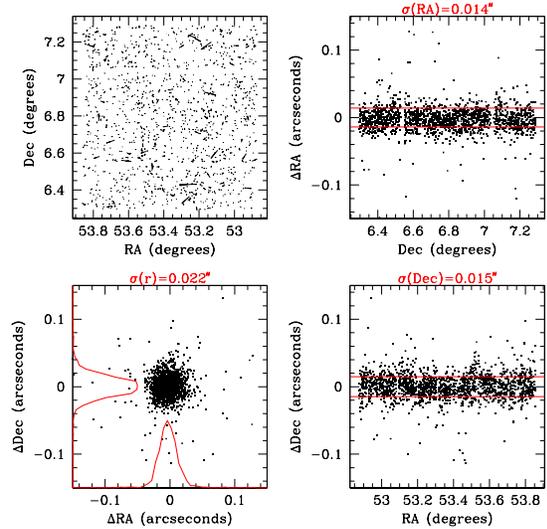} 
   \figcaption{Astrometric residuals between
two input images after matching to the ``merge-by-pixel'' master
catalog.  \textbf{Top left:}  the astrometric residuals as a
vector field. Lengths have been exaggerated for visual clarity. No patterns or trends in the residuals are apparent. \textbf{Bottom left:} residuals in RA and Dec. Histograms of the residuals in both directions are also
plotted. The title shows the standard deviation of the combined RA and Dec. squared sum of the residuals. \textbf{Right:} residuals in RA as a
function of Dec and residuals in Dec as a function of RA. The horizontal lines mark plus or minus 1 standard deviation from the mean of the residuals, the values of which are shown in the titles. \label{fig:occastrestyp}}
\end{figure}

Figure \ref{fig:occastrestyp} shows the astrometric residuals between
two input images after matching to the ``merge-by-pixel'' master
catalog.  The top left plot shows the astrometric residuals as a
vector field. The lengths have been greatly exaggerated. The bottom
left plot shows the residuals in RA and Dec as a scatter
plot. Histograms of the residuals in both directions are also
plotted. The two plots on the right show the residuals in RA as a
function of Dec and the residuals in Dec as a function of RA.  The
residuals are seen to be on the order of 0.02 arcseconds, while there
are a few mis-identifications (indicated by the outliers in the
scatter plots and by longer than usual lines in the vector field),
no large systematic shifts are apparent.

When the KBO of interest fell in one of the CFHT images, its position was measured with respect to the MPSC. For astrometry of the KBOs measured in the CFHT data, we adopt an uncertainty of 0.04", typical of the astrometric residuals found after the ``merge-by-pixel'' step, the last step in producing the MPSC. 

\subsection{Gemini Observations \label{sec:geminiObs}}
Along with the target astrometry provided by the MegaPrime data, additional astrometry of the targets was measured from images taken with the GMOS detector. Each target was visited multiple times. During each visit, a pair of images was taken with small $\sim30"$ dither between pairs. All exposures were taken in r'-filter and exposure times were tuned such that the resultant photometric signal-to-noise ratio (SNR) was $\sim40-50$. This ensured that the resultant astrometric error was dominated by the match between the images and the MPSC and not by the quality of the object measurement. The target's observed positions are presented in Table~\ref{tab:observations}.

After standard image reductions (bias removal and flat fielding) the astrometric plate solutions for the GMOS data were found by matching those images to the MPSC. The match was done with the use of SCAMP \citep{Bertin2006}. SCAMP is a program that produces astrometric solutions which match one point source catalog to a reference catalog. During the matching process, each image was treated individually, providing as  independent individual astrometric measurements as possible. The plate solution of GMOS was found to be adequately described by a simple CD matrix solution; utilizing higher order terms revealed no improvement in the astrometric solution. Experience has shown that SCAMP typically produces unreliable astrometric uncertainties. That is, the root mean square (RMS) astrometric scatter of a solution is typically erroneous in one of two ways. Either the reported RMS is small despite the fact that the astrometric solution is clearly erroneous, or the RMS value is found to be significantly smaller than the scatter in repeat images of the same targets. Repeated measurements of a KBO at a given epoch have demonstrated that the plate solutions provided by SCAMP result in a $\sim0.08"$ scatter in the astrometry of the tracked KBOs. Therefore, we forgo use of the quoted SCAMP RMS values, and adopt an astrometric uncertainty of $0.08"$ for all GMOS measurements. 

\subsection{Orbital Element Correction \label{sec:elem_corr}}
The second step in occultation predictions is the generation of accurate ephemerides. As discussed in Section~\ref{sec:Intro}, a common approach is what we refer to as the constant offset method, in which the nominal ephemerides of the targets in question are offset by a value representative of the typical differences between the observed and predicted positions of the object. 

One merit of the constant offset method comes as a result of the use of a nominal ephemeris for the target object. By use of a previously determined ephemeris, all past reported astrometry and an orbit determination from those data are automatically considered in the predictions. While it is certainly true that past astrometry are typically not of the quality required for occultation predictions, many measurements spanning a multi-year baseline provide modestly accurate ephemerides that, with small astrometric corrections, can be suitable for occultation predictions. 

\begin{figure}[h]
   \centering
   \plotone{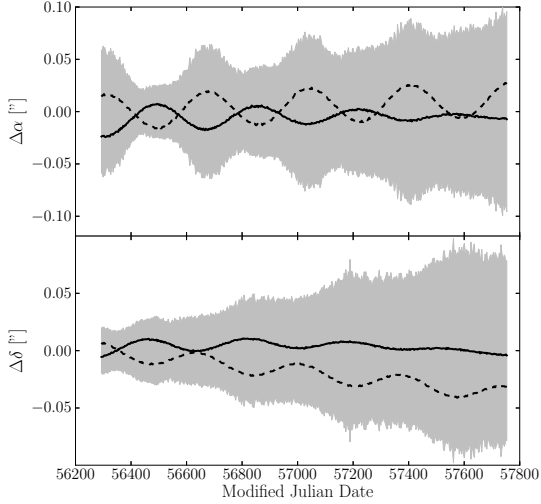} 
   \figcaption{Difference in Right Ascension (top) and Declination (bottom) as a function of MJD between the actual object ephemeris and that found by the constant offset approach (dashed) and the element correction method (solid) for a simulated observation of a KBO on orbit similar to Eris (see Section~\ref{sec:elem_corr}). The grey shaded region is the ephemeris envelope generated from the $1-\sigma$ range of orbital elements derived from the element correction likelihood routine. Note: the envelope only includes the range in ephemerides caused by the uncertain orbital elements, and does not include uncertainty in the position of the potentially occulted body.  \label{fig:ephem_comparison}}
\end{figure}

The accuracy of predictions made by the constant offset method primarily suffer from one major issue. In reality the difference between the nominal-offset ephemeris and the true ephemeris is not actually constant. The difference between the two ephemerides depends primarily on the quality of the nominal ephemeris. A small error in the nominal ephemeris can result in both a periodic shift (with period of a year) and a nearly linear shift between the nominal-offset ephemeris, and the true ephemeris (see Figure~\ref{fig:ephem_comparison}). Even for the best ephemerides, the amplitude of the periodic offset can be of order $\sim0.02"$. As a result, without frequent tracking and updates to the offset, predictions with this method can be unreliable. 

A different technique involves adjustment of the orbital elements of the nominal ephemeris to match available astrometry. That is, with appropriate tweaks to the nominal orbital elements, the ephemeris itself can be corrected and used directly for occultation predictions. We consider the orbital element adjustment approach and its comparison with the constant offset method here. 

We start with the orbital elements provided by the astDys catalog and make use of Orbfit\footnote{\url{http://adams.dm.unipi.it/orbfit/}} in calculating ephemerides. To determine the appropriate orbital element corrections, we adopt a maximum likelihood approach. Specifically, given adjustments to the semi-major axis, eccentricity, inclination, longitude of ascending node, argument of perihelion, and mean anomaly $\delta a, \delta e, \delta I, \delta \Omega, \delta \omega, \delta M$, we adopt the log-likelihood

\begin{equation}
\begin{split}
L(\delta a, \delta e, \delta I, \delta \Omega, \delta \omega, \delta M)=
P(\delta a, \delta e, \delta I, \delta \Omega, \delta \omega, \delta M) \\
-\frac{1}{2} \sum_{\textrm{i}} \left(  \frac{ \cos\beta_{\textrm{i}} \left(\alpha_{\textrm{i}}-\bar{\alpha_{\textrm{i}}}\right) }{\sigma_{\textrm{i},\alpha}}\right)^2+\left(\frac{ \beta_{\textrm{i}}-\bar{\beta_{\textrm{i}}} }{\sigma_{\textrm{i},\beta}}\right)^{2} 
\label{eq:likelihood}
\end{split}
\end{equation}

\noindent
where $\alpha_{\textrm{i}}$ and $\beta_{\textrm{i}}$ are the ith observed right ascension (RA) and declination (Dec) of the object, and $\bar{\alpha_{\textrm{i}}}$ and $\bar{\beta_{\textrm{i}}}$ are the RA and Dec predicted from the nominal ephemeris of the object in question. $\sigma_{\textrm{i},\alpha}$ and $\sigma_{\textrm{i},\beta}$ are the astrometric uncertainty of the ith measurement (recall that we adopt 0.04" and 0.08" uncertainty in the MegaPrime and GMOS observations respectively). 

In Equation~\ref{eq:likelihood}, $P(\delta a, \delta e, \delta I, \delta \Omega, \delta \omega, \delta M)$ is the log-prior on the orbital elements. For this, we turn to the element errors in the nominal ephemeris. We however, choose not to apply any priors on the orbital element angles. This decision is motivated by the consideration of the coordinate origin of the astrometric catalog (2MASS or SDSS) which we first consider when generating the MPSC. Each catalog will have a slightly different coordinate origin. The ephemerides provided by astDys reference a different zero point (usually that defined by the USNO A catalog). As a result of the differing origins,  the orbital angles $\Omega$, $\omega$, and $M$, are incorrect at the level of accuracy required. Similarly, as the nodal angle has changed, so has the orbital inclination. It is only the semi-major axis and eccentricity that are not affected by adjustment of the reference. We adopt gaussian priors on $a$ and $e$ with standard deviations equal to the uncertainties in those elements provided with the astDys ephemerides.

When determining the orbital element corrections with Equation~\ref{eq:likelihood}, we only consider the MegaPrime and GMOS observations we gathered ourselves and matched to the MPSC. These observations only span 1-2 years, and as a result partially avoid potential zonal errors which affect the astDys orbital elements. The disadvantage of this approach is that past observations are not directly used by our method. Rather, those data are only used as a prior, a choice which reflects how much prior information we can safely extract from the nominal astDys elements without producing unreliable results.

To converge on a maximum likelihood, we utilize MCMC Hammer \citep[EMCEE,][]{Foreman-Mackey2012}. MC Hammer is an affine invariant Markov Chain Monte-Carlo sampler which rapidly converges on the maximum likelihood solution in a minimum number of steps. For the routine, we adopt 100 walkers, and utilize a 100 step burn-in phase. During the maximization, EMCEE determines the autocorrelation time, $\tau$ - the number of steps required such that further samples are independent and distributed as the posterior likelihood function. Thus, the routine must be run for multiple $\tau$ to ensure that the likelihood space is accurately evaluated. For all orbital elements, $\tau$ was found to be $\sim5-8$. Thus, our routine was run for 150 steps after the burn-in phase, or roughly 20 or more autocorrelation times. Additional steps were found to not improve the resultant likelihood evaluations. 

As an end-product of this routine, we can extract confidence intervals on each of the orbital elements. From these confidence intervals, we can determine the astrometric uncertainty on the target for any future date and time. No direct method of positional astrometric uncertainty can be generated from the constant offset method.

To compare the performance of the constant offset and element adjustment methods, we set up a theoretical situation. For this example, we consider a theoretical Eris, and assigned as its elements the elements reported by astDys for the real Eris. Using Orbfit, we then simulated the historical observations of the theoretical KBO on the dates of all reported astrometry for the real Eris - the dates were extracted from the Minor Planet Center\footnote{\url{http://www.minorplanetcenter.org/iau/mpc.html}}. We note that Orbfit includes planetary perturbations, a necessary consideration for the decades long arcs of the KBOs we consider. Noise was added to the simulated observations to account for typical random uncertainties in reported astrometry. A gaussian distribution with standard deviation 0.3" in both RA and Dec was used. We also attempted to include zonal errors in the simulated observations. This was done by scattering the observations by a small value every time  the theoretical KBO moved more than 3 degrees; we adopted a gaussian distribution with width 0.08" in both RA and Dec. In a similar vein, we generated fake tracking observations (like those we present from MegaPrime and GMOS) adopting 0.04" astrometric uncertainties. 

For the theoretical KBO, a set of orbital parameters was determined from the simulated historical observations with Orbfit (ignoring the simulated tracking observations), the same routine used for the astDys elements. We then utilized our maximum likelihood approach to determine $(\delta a, \delta e, \delta I, \delta \Omega, \delta \omega, \delta M)$, the 6 element offsets which provided the best match to the simulated tracking observations. We also evaluated the best offset which minimized the residuals between the simulated tracking observations and the nominal ephemeris. 

The results of our simulation are shown in Figure~\ref{fig:ephem_comparison}, where we compare the results of both ephemeris correction processes. This figure clearly demonstrates one of the difficulties with occultation predictions. Both methods suffer from a 365 day period oscillation in the difference between the predicted and actual ephemerides. In addition, the constant offset method suffers from an increasing error in declination with time.

One advantage of the element correction approach is clear. The approach can produce more accurate ephemerides further away from the last epoch of observation than the constant offset method. The quality of our element correction approach however, depends critically on the quality of the astDys ephemeris. In addition, without sufficient baseline in the astrometry used in Equation~\ref{eq:likelihood}, the resultant element offsets can be incorrect; simulations suggest at the very least, 1-year baselines are required. The element correction approach is also sensitive to discrepant tracking astrometry which can result in ephemerides that quickly deviate away from the true ephemeris. Because the constant offset method averages all tracking astrometry, this method is much less sensitive to discrepancies in the tracking observations.

Which of the constant offset method or the element adjustment approach produces the most reliable occultation predictions will be different for each object, and the determination of which essentially requires future observations. Eventually, the element adjustment approach should surpass the constant offset method. How long is required is not easily determined a priori, but simulations suggest a $\sim2$~year baseline or longer will allow the element adjustment approach to surpass the constant offset method. Both methods however, suffer from different effects, and as a result, comparison of occultations predictions produced with both methods should be considered until the results of the element adjustment approach have been demonstrated to be superior on an object by object basis. Thus, we choose to present the results of both methods when reporting candidate occultations.

It should be noted that, along with the ephemeris uncertainty, there is also uncertainty in the stellar astrometry. From the data used in generating the MPSC, each star is observed multiple times, with a positional accuracy of $\sim 0.04"$ in both RA and Dec. The quoted stellar position represents a global mean of each measurement. As we will discuss further below, it appears that the uncertainty on the mean position does not decrease as the square-root of the number of times the star is observed, but rather, appears roughly a factor of 2 worse than that expectation. 


\subsection{Proper Motions and Chromatic Differential Refraction}

During the generation of the MPSC, we make no effort to account for the effects or proper motion, or chromatic differential refraction. The first effect, that of proper motions, was ignored during our analysis because proper motions were not determined for either of the reference catalogs we used. As a result, the stars used in the first step of the astrometric calibration will have moved slightly since they were observed by 2MASS (1997-2001) or the SDSS (2000 to present). Estimates of the proper motion of sources in  the 2MASS catalog are available from the PPMXL catalog \citep{Roeser2010}. Experimentation with that catalog revealed that the proper motions measured by PPMXL were only accurate enough to improve the positions of the brightest sources in the 2MASS catalog, those which were primarily saturated in our observations. No noticeable improvement in positional accuracy of fainter sources from which our astrometric solutions were derived, was found. As a result, we chose to adopt the cataloged positions for the first stage of our astrometric calibrations and make no effort to correct for proper motions.

\citet{Tholen2013} point out that not considering proper motions can result in zonal errors. In the small fields we observe to produce the MPSC, these zonal errors would manifest themselves as a nearly constant systematic offset between the frame of the refence catalog (2MASS or SDSS) and the frame of the MPSC. The offset will be nearly constant across the small fields covered by the MPSC for each object. 

To determine if ignoring proper motions could result in inaccurate predictions, we performed a test of the predictions for the object 2002 TC302, in which the positions of stars in the reference catalog (2MASS) were scattered to mimic the effects of proper motions. The distribution of proper motions was extracted from the PPMXL catalog in a 2 degree wide patch around the position of the KBO. The RA and Dec positions of each source were then randomly scattered according to the distribution of RA and Dec proper motions. Analysis of the PPMXL proper motions suggest that $\sim15\%$ of sources have moved more than 1" since the 2MASS catalog was created, and hence would be ignored during the MPSC generation. After the sources were randomly scattered, this number increased by a factor of $\sim\sqrt2$. The full occultation predictions routine (MPSC generation, determination of KBO astrometry, and ephemeris correction) was then applied using the randomized reference catalog. As predicted, compared to the astrometry referenced to the non-scattered 2MASS catalog, a systematic offset in both the MPSC source positions and that of the KBO were found, with amplitude of nearly 0.1" in RA, and 0.05" in Dec. In addition, the astrometry from the scattered catalog had a slope of $\sim0.05"$ per degree RA in both the  RA and Dec axes. The effects of the offset and slope produced a shift of $\sim300$~km in the predicted shadow tracks of 2002 TC302. Even the most accurate predictions have $\sim1000$~km uncertainties in the shadow path which is entirely a result of ephemeris uncertainty. As ephemeris uncertainty grows rapidly with time away from the last tracking observation, the prediction errors induced by not accounting for proper motions will always be small compared to the ephemeris error. 

The reason why not accounting for proper motions has such little effect on the predictions can be easily understood. Recall that during the MPSC generation, an external catalog is only used in the first stage of generating astrometric solutions. An internal catalog is used in later stages to refine those solutions and generate a self-consistent MPSC. As a result of this internal reference to the observed sources, small variations in the solution caused by proper motions are reduced at each subsequent iteration of our calibration routine, mitigating these issues. In general, as shown by our test with 2002 TC302, this systematic offset does not significantly affect the quality of the astrometric solution or the KBO ephemeris as the astrometry of both the MPSC and the KBO are equally affected by the offsets. 

While proper motions in general do not affect the quality of the astrometric solution, they will cause a degradation of some individual predictions with time away from the observations used to generate the MPSC. Take for example, a star with proper motion of 0.1" per year - roughly 15\% of all stars have at least this proper motion. At the time of observation of that star for MPSC generation, its position is known to the precision of the MPSC. At typical KBO distances, the high proper motion of the star will have moved the shadow track more than 1000~km from that predicted in just 6 months. Clearly, advanced monitoring of the target stars is warranted for particularly profitable events.

Like proper motions, we make no effort to account for the effects of chromatic differential refraction (CDR), primarily due to the single band observations we acquired for MPSC generation. Our observations were taken over a broad range of airmasses, but less than 2 in all cases. In the standard conditions on Mauna Kea \citep{Cohen1988}, the reddest stars, with (V-I)$\sim3$, will experience a shift of amplitude $\sim0.05"$ at an airmass of 2 compared to its relative position when observed at zenith \citep{Stone1984,Monet1992}. For most stars, (V-I)$\approx0.8$ resulting in a shift is of order 0.01". The reddest KBOs eg. Quaoar and Sedna with (V-I)$\sim1.3$, will experience a slightly larger shift. The primary consequence of CDR is a fundamental limit on the precision of the MPSC; unless observations of the MPSC sources are taken at random airmasses, their astrometry cannot be more precise than the shift cause by CDR at their average observed airmass. For most stars, and our observations, we conservatively estimate the amplitude of this effect at up to a $\sim0.02"$ shift of individual stars compared to the astrometry of the KBO. Improvements beyond this limit cannot be made, unless either the effects of CDR are accounted for, requiring multi-band observations, or observations are restricted to a small range in airmasses. 

\section{The Predictions \label{sec:predictions}}

\begin{figure}[h]
   \centering
   \plotone{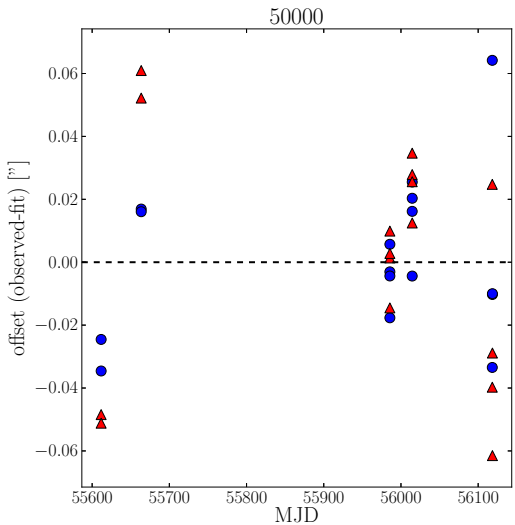} 
   \figcaption{Astrometric residuals in arcseconds as a function of date of observation after application of the element correction approach for (50000) Quaoar. Residuals in RA and Dec are shown as blue circles and red triangles respectively.   \label{fig:50000}}
\end{figure}

\begin{figure}[h]
   \centering
   \plotone{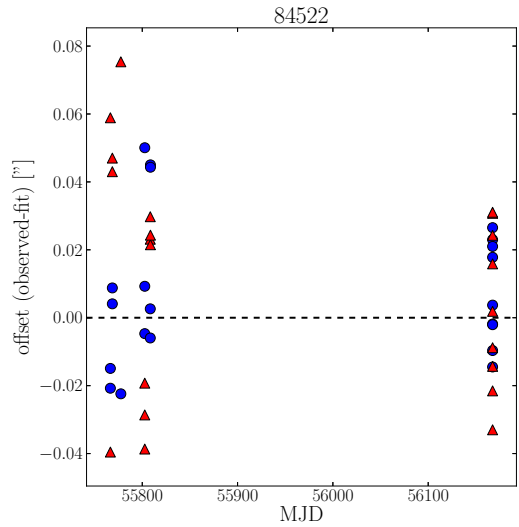} 
   \figcaption{As in Figure~\ref{fig:50000} but for object (84522) 2002 TC302.   \label{fig:84522}}
\end{figure}

\begin{figure}[h]
   \centering
   \plotone{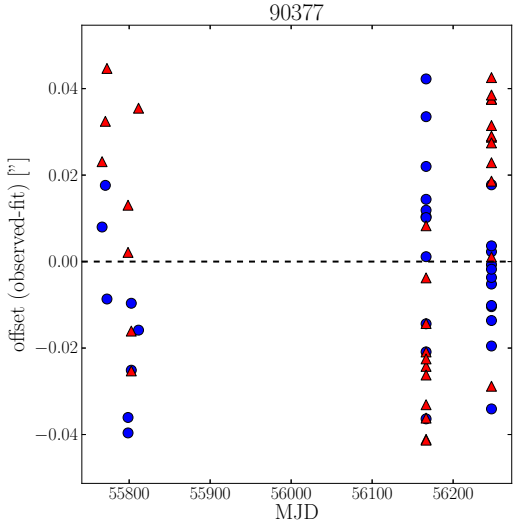} 
   \figcaption{As in Figure~\ref{fig:50000} but for object (90377) Sedna.   \label{fig:90377}}
\end{figure}

\begin{figure}[h]
   \centering
   \plotone{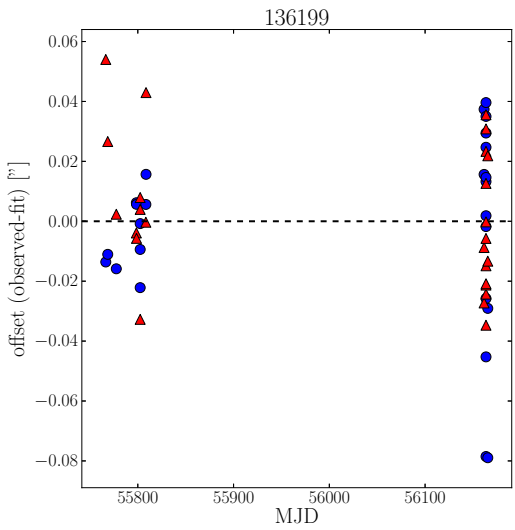} 
   \figcaption{As in Figure~\ref{fig:50000} but for object (136199) Eris.   \label{fig:136199}}
\end{figure}

\begin{figure}[h]
   \centering
   \plotone{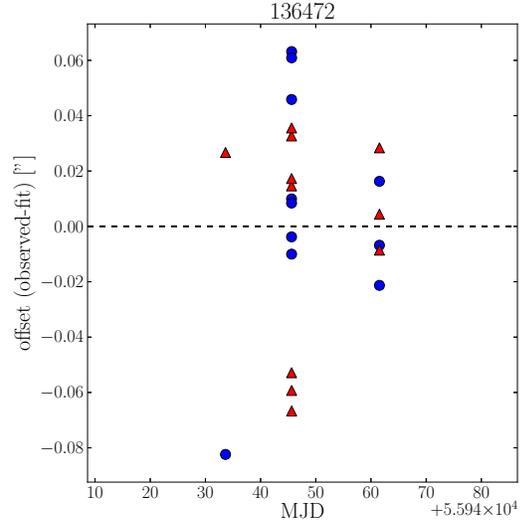} 
   \figcaption{As in Figure~\ref{fig:50000} but for object (136472) Makemake.   \label{fig:136472}}
\end{figure}

\begin{figure}[h]
   \centering
   \plotone{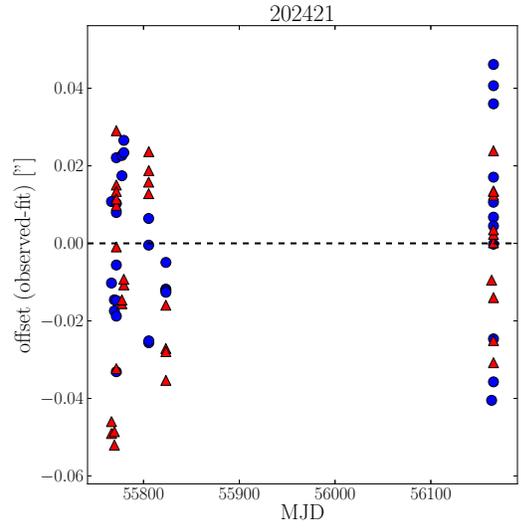} 
   \figcaption{As in Figure~\ref{fig:50000} but for object (202421) 2005 UQ513.   \label{fig:202421}}
\end{figure}

\begin{figure}[h]
   \centering
   \plotone{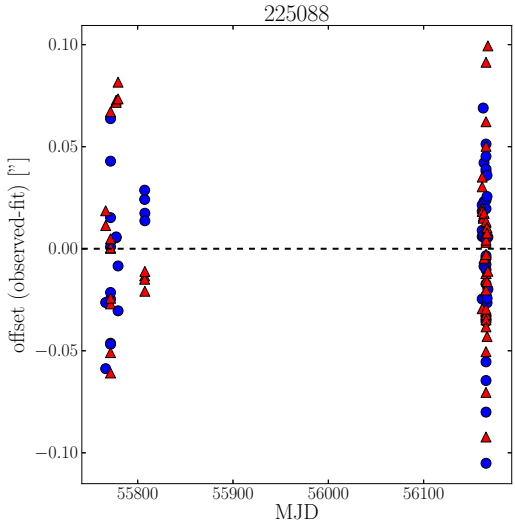} 
   \figcaption{As in Figure~\ref{fig:50000} but for object (225088) 2007 OR10.   \label{fig:225088}}
\end{figure}

Both the constant offset method and our element correction routine were used to produce occultation predictions for the 7 targets of this study. The resultant best-fit orbital elements - taken as the MCMC point with highest likelihood - are presented in Table~\ref{tab:elements}. All available occultations are presented at \url{www.fraserkbos.com}. A few notable predictions are presented in Table~\ref{tab:predictions}. The residuals of the element correction approach are presented in Figures ~\ref{fig:50000}, \ref{fig:84522}, \ref{fig:90377}, \ref{fig:136199}, \ref{fig:136472}, \ref{fig:202421}, and \ref{fig:225088}.

It is important to note that the coverage of the future tracks by our MPSC is different for each object. For instance, for Eris, the occultations can be predicted as far into the future as January 1, 2015. For Sedna however, this is only June 1, 2014. Efforts will be made to ensure future coverage will be made available. As a practical limit, we only report candidate occultations by stars with magnitude brighter than r'=21.

\begin{figure}[h]
   \centering
   \plotone{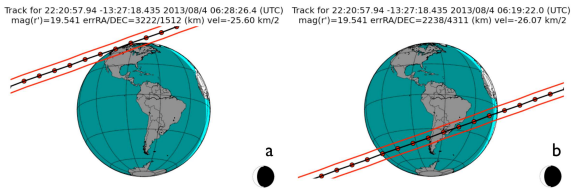} 
   \figcaption{Example prediction for occultation by object 2007 OR10. Predictions using the element correction approach and constant offset method are shown in \textbf{a} and \textbf{b} respectively. Properties of the event, including star position and brightness, nominal occultation center time and velocity, and an estimate of the uncertainty are presented. Nominal shadow extent shown with red lines. Red dots are spaced 1 minute apart. Day-night terminator and moon phase are shown at the nominal occultation center time. Positional uncertainty of each method is shown. OR10 is assumed to be 1200~km in diameter.   \label{fig:pred}}
\end{figure}

An example of a candidate occultation by 2007 OR10 is shown in Figure~\ref{fig:pred}. The candidate star has r'=19.5; and  with predicted time roughly 2013 August 8 6:20 UT. The uncertainties presented in the prediction depend on the method. The uncertainties quoted by the element correction routine are the 1-$\sigma$ scatter in RA and Dec produced by the MCMC fit, and for the constant offset method are just the RMS residuals left in the astrometry of 2007 OR10.  This candidate demonstrates an extreme case for the difference in predictions possible between the two methods, roughly 4000~km. The element adjustment approach suggests that the large discrepancy is caused by the oscillations between the true and predicted positions inherent to the constant offset method; a large offset of $\sim0.05"$ occurs on this date. Only with observations near this event will this be confirmed. 

\section{Confirmation \label{sec:confirmation}}
An occultation by the KBO Sedna was predicted to be visible over North America at 03:40, Dec 26 2012 UT of a star with r'=18.9.  Attempts  to observe this event were made at various telescopes, including: the Plaskett telescope at the Dominion Astrophysical Observatory in Victoria, BC (WCF); the University of Wyoming Infrared Telescope, near Laramie WO (LAY); The R. A. Cross Telescope near Calgary, BC (PPL); the Las Cumbres Observatory Global Telescope at McDonald Observatory (TL and FBB); the Astrophysical Research Consortium Telescope at Apache Point (RJM); the Vatican Advanced Technology Telescope at Mount Graham (RPB); and the Perkins Telescope at Lowell Observatory (MJB). Unfortunately, a widespread storm system prevented useful data from being collected from any site.

Some confirmation of the accuracy of the predictions has been made possible from the observation of 3 candidate occultations of the object Quaoar made by the Gemini telescopes. Two candidates were observed on July 6 and 13, 2012 by the Gemini-South telescope of stars with r'=20.0 and 20.2 respectively. The third candidate was of a r'=16.37 magnitude star observed by the Gemini-North telescope on July 10, 2012. Observations were made with the GMOS cameras roughly 1--2 hours before and after the nominal event times; pairs of images with 40~s exposure times in the r' filter were taken. These allowed us to accurately measure the occultation impact parameter of each event. The events themselves were observed with the Acquisition Cameras in the R-filter in windows approximately 20 minutes in length centred on the nominal event times. These observations will be reported in a separate manuscript. Weather frustrated the observations on July 6; no useful data was acquired on that date.

\begin{figure}[h]
   \centering
   \plotone{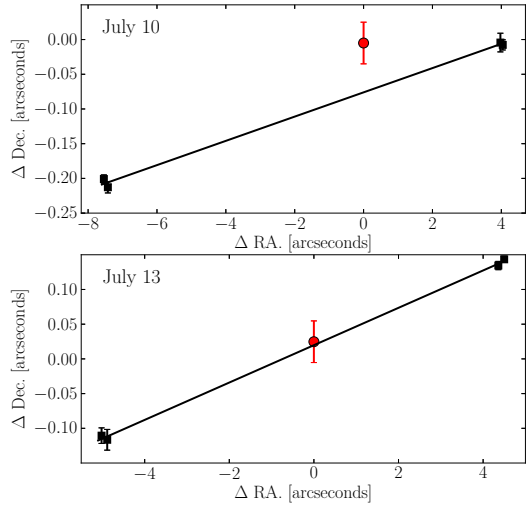} 
   \figcaption{Diagrams showing distance of closes approach for the July 10 (top) and 13 (bottom) events. Measured offsets (Quaoar-Star) are shown as black squares. Quaoar's inferred trajectory is shown as the black line. The Closest approach predicted by the element adjustment approach  is shown as the red circle. The uncertainty on the predicted impact parameter only includes that derived from the element correction approach, and does not include the uncertainty in target star position. Note: uncertainties in RA. are smaller than the data points in this figure, but are the same size as the Dec. uncertainties.  \label{fig:impacts}}
\end{figure}

Predicted offsets between Quaoar and the target star at the time of GMOS observations for the July 10 and 13 events are presented in Table~\ref{tab:offsets}. Observed offsets were measured using the IRAF {\it daophot} gaussian centroid routine \citep{Tody1993}. The observed offset between Quaoar and the target star in each image are reported in Table~\ref{tab:offsets}. Using these measured offsets, Quaoar's impact parameter for each event was determined by fitting a straight line to the observed offsets and finding the closest point between the line and star position. Uncertainties on the impact parameters were then found by a Monte Carlo approach. For each of the four offset measurements, a random point was generated within the measurement uncertainties, and the impact parameter of that random realization was found. This process was repeated to 1000 times to generate a range of impact parameters consistent with the observations. We quote the standard deviation of these realizations as the uncertainty on the event impact parameters. Diagrams presenting Quaoar's trajectory with respect to the target stars are shown in Figure~\ref{fig:impacts}.

The July 13 event was predicted to have an impact parameter of $0.017\pm0.03"$ and $0.0"$ from the element adjustment and constant offset methods respectively. The observed impact parameter was $0.019\pm0.004"$, in excellent agreement with the element adjustment approach. 

The July 10 event was predicted to have an impact parameter of $0.012\pm0.03"$ and $0.03"$ from the element adjustment and constant offset methods respectively. The measured impact parameter was $0.076\pm0.005"$. This result may be interpreted that for this event, the constant offset method produced a more accurate prediction. Given the short 3 day interval between the events on July 10 and 13, the difference between the true ephemeris and those produced by the constant offset method are virtually the same on both days. As a result, each method must produce predictions of similar quality on July 10 as they produced for July 13. It must be that the position of the July 10 target star was not sufficiently well known. 

The July 10 target star has been tentatively identified with a star in the USNO catalog with colour (R-I)$\approx0.4$. Comparison with Quaoar, (R-I)=1.3, suggests that CDR may result in a shift of as much as $\sim0.015"$; CDR cannot account for the observed discrepancy in the target star's position with respect to Quaoar. It may be that the star had a high proper motion. To account for the discrepancy, the star would need to have a proper motion at least 0.2"/year. The fraction of stars with proper motion at least that high is only 10\%. Thus, it seems unlikely that proper motions are the cause of the discrepancy. Recall that the adopted uncertainty in position from the MPSC is 0.04". Scaling this value by the square root of the number of MegaPrime observations of the star, 6, the expected astrometric uncertainty on the star is $\sim0.016"$. If this were the true stellar position uncertainty, then the observed impact parameter represents a more than 2-$\sigma$ deviation from the prediction; it must be that the stellar astrometric uncertainty does not decrease as rapidly as $\sqrt{N}$. This suggests that in the MPSC generation, each individual measurement of a star is not fully independent, but rather the measurements of a star are partially correlated.  The true stellar astrometric uncertainty seems as much as a factor of $\sim2$ larger than that expectation.

That the observed impact parameter of the July 13 event was in agreement with predictions demonstrates the utility of the element adjustment method in accurately predicting occultations. The observations reinforce the findings of our simulations. The element adjustment approach can be used to accurately predict stellar occultations. Until this approach has been shown to be superior for a particular object, other methods should also be considered alongside the element adjustment approach to gauge the uncertainty in a particular event. Further, our findings demonstrate the importance of knowing the stellar position, which can be as large as the uncertainty in the ephemerides. Our findings suggest that $\sim30$ individual observations of a star will be required before its position is known to better than $0.01"$. This suggests that stellar position uncertainty can be the dominant factor in overall prediction uncertainty.

\section{Conclusions}
We have developed a new method of occultation predictions by which the cataloged orbital elements of Kuiper Belt Objects are corrected for the difference between the ephemerides predicted by those elements, and the observed ephemerides. Observations of the fields occupied by 7 well-tracked KBOs were acquired. From those observations, extremely accurate master point source catalogs were generated. We applied the element correction method as well as the standard constant offset method to the observations of the KBOs, and generated occultation predictions from both methods. The results of both methods were compared and it was found that the constant offset method suffers primarily from inaccurate cataloged ephemerides. We found that the element correction method suffers more from inaccuracies in the observations used to correct the orbital elements. For well tracked objects however, the element correction method seems to produce corrected ephemerides that degrade much more slowly with time than does the constant offset approach.
\\

\acknowledgements
Based on observations obtained with MegaPrime/MegaCam, a joint project of CFHT and CEA/IRFU, at the Canada-France-Hawaii Telescope (CFHT) which is operated by the National Research Council (NRC) of Canada, the Institut National des Science de l'Univers of the Centre National de la Recherche Scientifique (CNRS) of France, and the University of Hawaii. This work is based in part on data products produced at Terapix available at the Canadian Astronomy Data Centre as part of the Canada-France-Hawaii Telescope Legacy Survey, a collaborative project of NRC and CNRS

Based on observations obtained at the Gemini Observatory, which is operated by the 
Association of Universities for Research in Astronomy, Inc., under a cooperative agreement 
with the NSF on behalf of the Gemini partnership: the National Science Foundation 
(United States), the National Research Council (Canada), CONICYT (Chile), the Australian 
Research Council (Australia), Minist\'{e}rio da Ci\^{e}ncia, Tecnologia e Inova\c{c}\~{a}o 
(Brazil) and Ministerio de Ciencia, Tecnolog\'{i}a e Innovaci\'{o}n Productiva (Argentina).

Federica B Bianco is supported at NYU by a James Arthur fellowship.


\begin{deluxetable}{llllllll} 
	\tabletypesize{\small}
	\tablecaption{Approximate CFHT MegaPrime Sky Coverage \tablenotemark{a}\label{tab:CFHT_coverage}}
	\tablehead{
	\colhead{Object}  & \colhead{$\alpha_{\textrm{min}}$ (deg)} & \colhead{$\alpha_{\textrm{max}}$ (deg)} & \colhead{$\delta_{\textrm{min}}$ (deg)} & \colhead{$\delta_{\textrm{max}}$ (deg)} & \colhead{Latest Predictions\tablenotemark{b}} & \colhead{External Reference Catalog}
	}
	\startdata
	(50000) Quaoar & 261.73 & 264.27 & -15.75 & -15.36 & 2013 & 2MASS\\
	(84522) 2002 TC302 & 30.54 & 22.66 & 22.26 & 24.15 & 2013 & 2MASS\\
	(90377) Sedna & 52.09 & 54.50 & 6.48 & 7.36 & 2015 & SDSS\\
	(136199) Eris & 24.40 & 25.99 & -4.35 & -2.86 & 2015 & SDSS\\
	(136472) Makemake & 189.26 & 191.82 & 26.92 & 27.81 & 2014 & SDSS\\
	(202421) 2005 UQ513 & 4.24 & 7.46 & 28.69 & 30.76 & 2013 & SDSS\\
	(225088) 2007 OR10 & 333.92 & 335.29 & -14.38 & -13.05 & 2015 & 2MASS\\
	\enddata
	\tablenotetext{a}{- Coordinates are in the J2000 reference frame.}
	\tablenotetext{b}{- at least partial coverage of the object's ephemeris during this year.}
\end{deluxetable}

\begin{deluxetable}{llll} 
	\tabletypesize{\small}
	\tablecaption{Object Astrometry \label{tab:observations}}
	\tablehead{
	\colhead{MJD}  & \colhead{$\alpha$ (deg)\tablenotemark{a}} & \colhead{$\delta$ (deg)\tablenotemark{a}} & \colhead{Coordinate Uncertainty (")}
	}
	\startdata
\multicolumn{4}{l}{(50000) Quaoar} \\ \hline
55611.655578 & 262.317292 & -15.682684 & 0.08 \\
55611.657210 & 262.317310 & -15.682681 & 0.08 \\
55663.553265 & 262.453589 & -15.529130 & 0.08 \\
55663.554898 & 262.453576 & -15.529122 & 0.08 \\
55985.641534 & 263.731585 & -15.674959 & 0.04 \\
55985.642611 & 263.731593 & -15.674956 & 0.04 \\
55985.643692 & 263.731603 & -15.674958 & 0.04 \\
55985.644775 & 263.731610 & -15.674948 & 0.04 \\
56014.560038 & 263.854030 & -15.588429 & 0.04 \\
56014.561118 & 263.854034 & -15.588423 & 0.04 \\
56014.589606 & 263.853980 & -15.588339 & 0.04 \\
56014.590690 & 263.853977 & -15.588332 & 0.04 \\
56118.285567 & 262.292690 & -15.379824 & 0.08 \\
56118.286968 & 262.292638 & -15.379842 & 0.08 \\
56118.464271 & 262.289430 & -15.379901 & 0.08 \\
56118.465683 & 262.289404 & -15.379892 & 0.08 \\
\hline \\
\multicolumn{4}{l}{(84522) 2002 TC302} \\ \hline
55766.549223 & 32.684558 & 22.716384 & 0.08 \\
55766.552130 & 32.684566 & 22.716376 & 0.08 \\
55768.610035 & 32.690913 & 22.729657 & 0.08 \\
55768.612951 & 32.690921 & 22.729677 & 0.08 \\
55777.535649 & 32.700664 & 22.781575 & 0.08 \\
55802.626603 & 32.575117 & 22.874536 & 0.04 \\
55802.627639 & 32.575122 & 22.874536 & 0.04 \\
55802.628671 & 32.575101 & 22.874535 & 0.04 \\
55808.413270 & 32.515826 & 22.884205 & 0.08 \\
55808.416170 & 32.515795 & 22.884211 & 0.08 \\
55808.587584 & 32.513837 & 22.884434 & 0.08 \\
55808.590500 & 32.513803 & 22.884438 & 0.08 \\
56167.565781 & 33.379745 & 23.861149 & 0.04 \\
56167.566890 & 33.379726 & 23.861162 & 0.04 \\
56167.567966 & 33.379719 & 23.861156 & 0.04 \\
56167.569045 & 33.379709 & 23.861170 & 0.04 \\
56167.570125 & 33.379706 & 23.861155 & 0.04 \\
56167.571205 & 33.379695 & 23.861175 & 0.04 \\
56167.572363 & 33.379676 & 23.861170 & 0.04 \\
56167.573443 & 33.379676 & 23.861166 & 0.04 \\
56167.574524 & 33.379659 & 23.861179 & 0.04 \\
\hline \\
\multicolumn{4}{l}{(90377) Sedna} \\ \hline
55766.573408 & 53.498839 & 6.991800 & 0.08 \\
55770.605386 & 53.521642 & 6.988467 & 0.08 \\
55772.597075 & 53.531882 & 6.986487 & 0.04 \\
55798.616249 & 53.600415 & 6.942037 & 0.08 \\
55798.617406 & 53.600414 & 6.942031 & 0.08 \\
55802.630832 & 53.599887 & 6.932450 & 0.04 \\
55802.631862 & 53.599891 & 6.932445 & 0.04 \\
55811.575325 & 53.588023 & 6.909029 & 0.08 \\
56166.600171 & 54.159237 & 7.070095 & 0.04 \\
56166.602789 & 54.159258 & 7.070093 & 0.04 \\
56166.603917 & 54.159248 & 7.070082 & 0.04 \\
56166.605121 & 54.159248 & 7.070078 & 0.04 \\
56166.607849 & 54.159247 & 7.070067 & 0.04 \\
56166.610168 & 54.159244 & 7.070065 & 0.04 \\
56166.611296 & 54.159237 & 7.070061 & 0.04 \\
56166.612487 & 54.159239 & 7.070061 & 0.04 \\
56166.613652 & 54.159246 & 7.070053 & 0.04 \\
56166.617210 & 54.159250 & 7.070052 & 0.04 \\
56166.618394 & 54.159247 & 7.070043 & 0.04 \\
56247.334949 & 53.655513 & 6.833374 & 0.04 \\
56247.336107 & 53.655502 & 6.833367 & 0.04 \\
56247.337203 & 53.655493 & 6.833367 & 0.04 \\
56247.338359 & 53.655486 & 6.833362 & 0.04 \\
56247.339439 & 53.655467 & 6.833361 & 0.04 \\
56247.340518 & 53.655459 & 6.833361 & 0.04 \\
56247.341874 & 53.655446 & 6.833359 & 0.04 \\
56247.343063 & 53.655432 & 6.833337 & 0.04 \\
56247.344141 & 53.655424 & 6.833354 & 0.04 \\
56247.345223 & 53.655403 & 6.833348 & 0.04 \\
56247.346303 & 53.655403 & 6.833338 & 0.04 \\
56247.347381 & 53.655386 & 6.833342 & 0.04 \\
\hline \\
\multicolumn{4}{l}{(136199) Eris} \\ \hline
55766.556284 & 25.602261 & -3.768848 & 0.08 \\
55768.604971 & 25.601333 & -3.773118 & 0.08 \\
55777.540434 & 25.589507 & -3.794366 & 0.08 \\
55798.486896 & 25.514069 & -3.857934 & 0.08 \\
55798.630434 & 25.513314 & -3.858422 & 0.08 \\
55802.480507 & 25.492609 & -3.871678 & 0.04 \\
55802.481538 & 25.492597 & -3.871693 & 0.04 \\
55802.482653 & 25.492594 & -3.871686 & 0.04 \\
55808.409195 & 25.456987 & -3.892695 & 0.08 \\
55808.580605 & 25.455872 & -3.893299 & 0.08 \\
56161.606367 & 25.652157 & -3.591412 & 0.04 \\
56161.607444 & 25.652158 & -3.591421 & 0.04 \\
56163.593624 & 25.642338 & -3.598028 & 0.04 \\
56163.594707 & 25.642339 & -3.598037 & 0.04 \\
56163.595783 & 25.642330 & -3.598044 & 0.04 \\
56163.596861 & 25.642332 & -3.598044 & 0.04 \\
56163.597941 & 25.642308 & -3.598038 & 0.04 \\
56163.600103 & 25.642292 & -3.598039 & 0.04 \\
56163.604420 & 25.642286 & -3.598057 & 0.04 \\
56163.608742 & 25.642269 & -3.598078 & 0.04 \\
56163.609826 & 25.642262 & -3.598085 & 0.04 \\
56163.619655 & 25.642180 & -3.598105 & 0.04 \\
56163.622955 & 25.642186 & -3.598131 & 0.04 \\
56165.465516 & 25.632600 & -3.604358 & 0.04 \\
56165.466591 & 25.632608 & -3.604352 & 0.04 \\
\hline \\
\multicolumn{4}{l}{(136472) Makemake} \\ \hline
55973.644682 & 190.974085 & 27.297554 & 0.04 \\
55985.608243 & 190.816955 & 27.419601 & 0.04 \\
55985.609359 & 190.816934 & 27.419614 & 0.04 \\
55985.612684 & 190.816894 & 27.419645 & 0.04 \\
55985.616096 & 190.816843 & 27.419663 & 0.04 \\
55985.617177 & 190.816818 & 27.419690 & 0.04 \\
55985.624423 & 190.816726 & 27.419742 & 0.04 \\
55985.625504 & 190.816708 & 27.419747 & 0.04 \\
56001.535272 & 190.566893 & 27.561053 & 0.04 \\
56001.536352 & 190.566867 & 27.561055 & 0.04 \\
56001.537443 & 190.566844 & 27.561069 & 0.04 \\
56030.539757 & 190.064060 & 27.720943 & 0.04 \\
56030.540836 & 190.064051 & 27.720934 & 0.04 \\
56030.543005 & 190.064007 & 27.720949 & 0.04 \\
56326.479428 & 191.949982 & 26.705260 & 0.04 \\
56326.480505 & 191.949975 & 26.705269 & 0.04 \\
56326.481599 & 191.949959 & 26.705275 & 0.04 \\
56326.482708 & 191.949958 & 26.705285 & 0.04 \\
56326.483785 & 191.949948 & 26.705309 & 0.04 \\
56326.488112 & 191.949909 & 26.705350 & 0.04 \\
\hline \\
\multicolumn{4}{l}{(202421) 2005 UQ513} \\ \hline
55766.542406 & 6.328121 & 29.393568 & 0.15 \\
55766.545329 & 6.328108 & 29.393587 & 0.15 \\
55769.575632 & 6.307870 & 29.413344 & 0.15 \\
55769.579379 & 6.307842 & 29.413369 & 0.15 \\
55771.604166 & 6.292654 & 29.425701 & 0.04 \\
55771.605415 & 6.292632 & 29.425695 & 0.04 \\
55771.606501 & 6.292635 & 29.425715 & 0.04 \\
55771.607663 & 6.292621 & 29.425726 & 0.04 \\
55771.608740 & 6.292620 & 29.425727 & 0.04 \\
55771.609935 & 6.292599 & 29.425734 & 0.04 \\
55771.611162 & 6.292590 & 29.425738 & 0.04 \\
55777.429249 & 6.241761 & 29.457017 & 0.15 \\
55777.432170 & 6.241731 & 29.457031 & 0.15 \\
55779.429720 & 6.221865 & 29.466342 & 0.15 \\
55779.432634 & 6.221833 & 29.466356 & 0.15 \\
55805.378961 & 5.868124 & 29.517194 & 0.15 \\
55805.381869 & 5.868073 & 29.517191 & 0.15 \\
55805.529037 & 5.865584 & 29.517108 & 0.15 \\
55805.531951 & 5.865535 & 29.517104 & 0.15 \\
55823.328227 & 5.548573 & 29.474693 & 0.15 \\
55823.331638 & 5.548508 & 29.474683 & 0.15 \\
55823.385840 & 5.547476 & 29.474463 & 0.15 \\
55823.388756 & 5.547422 & 29.474449 & 0.15 \\
56163.513767 & 6.946110 & 30.082751 & 0.04 \\
56165.452640 & 6.916665 & 30.084417 & 0.04 \\
56165.453721 & 6.916659 & 30.084404 & 0.04 \\
56165.454800 & 6.916640 & 30.084412 & 0.04 \\
56165.455882 & 6.916628 & 30.084416 & 0.04 \\
56165.456987 & 6.916619 & 30.084414 & 0.04 \\
56165.458064 & 6.916592 & 30.084418 & 0.04 \\
56165.459158 & 6.916565 & 30.084416 & 0.04 \\
56165.460234 & 6.916565 & 30.084412 & 0.04 \\
56165.461314 & 6.916549 & 30.084408 & 0.04 \\
56165.469550 & 6.916410 & 30.084426 & 0.04 \\
\hline \\
\multicolumn{4}{l}{(225088) 2007 OR10} \\ \hline
55766.416929 & 334.910094 & -13.958308 & 0.08 \\
55766.422054 & 334.910058 & -13.958326 & 0.08 \\
55771.515914 & 334.864568 & -13.974438 & 0.04 \\
55771.516976 & 334.864570 & -13.974449 & 0.04 \\
55771.518146 & 334.864541 & -13.974452 & 0.04 \\
55771.519315 & 334.864544 & -13.974448 & 0.04 \\
55771.520406 & 334.864551 & -13.974467 & 0.04 \\
55771.521585 & 334.864509 & -13.974438 & 0.04 \\
55771.523067 & 334.864503 & -13.974461 & 0.04 \\
55771.524242 & 334.864509 & -13.974482 & 0.04 \\
55777.419374 & 334.809116 & -13.993759 & 0.08 \\
55777.423444 & 334.809076 & -13.993773 & 0.08 \\
55779.417995 & 334.789760 & -14.000428 & 0.08 \\
55779.422065 & 334.789714 & -14.000439 & 0.08 \\
55807.314056 & 334.504747 & -14.094050 & 0.08 \\
55807.318130 & 334.504705 & -14.094063 & 0.08 \\
55807.530686 & 334.502484 & -14.094744 & 0.08 \\
55807.534751 & 334.502441 & -14.094755 & 0.08 \\
56161.394827 & 334.840370 & -13.786288 & 0.04 \\
56161.395965 & 334.840359 & -13.786293 & 0.04 \\
56161.397133 & 334.840337 & -13.786293 & 0.04 \\
56161.398291 & 334.840337 & -13.786296 & 0.04 \\
56161.399438 & 334.840326 & -13.786317 & 0.04 \\
56162.468039 & 334.829289 & -13.789909 & 0.04 \\
56162.469176 & 334.829260 & -13.789913 & 0.04 \\
56163.325587 & 334.820436 & -13.792798 & 0.04 \\
56163.326731 & 334.820429 & -13.792796 & 0.04 \\
56163.327896 & 334.820403 & -13.792800 & 0.04 \\
56165.369806 & 334.799285 & -13.799639 & 0.04 \\
56165.370963 & 334.799272 & -13.799655 & 0.04 \\
56165.372043 & 334.799253 & -13.799670 & 0.04 \\
56165.408962 & 334.798865 & -13.799799 & 0.04 \\
56165.410042 & 334.798870 & -13.799776 & 0.04 \\
56165.411124 & 334.798834 & -13.799786 & 0.04 \\
56165.412370 & 334.798838 & -13.799788 & 0.04 \\
56165.413446 & 334.798833 & -13.799789 & 0.04 \\
56165.414525 & 334.798829 & -13.799789 & 0.04 \\
56165.415866 & 334.798780 & -13.799811 & 0.04 \\
56165.416946 & 334.798782 & -13.799783 & 0.04 \\
56165.418031 & 334.798781 & -13.799808 & 0.04 \\
56165.419294 & 334.798737 & -13.799794 & 0.04 \\
56165.420371 & 334.798766 & -13.799821 & 0.04 \\
56165.421605 & 334.798737 & -13.799824 & 0.04 \\
56165.422682 & 334.798742 & -13.799794 & 0.04 \\
56165.423765 & 334.798713 & -13.799820 & 0.04 \\
56165.425020 & 334.798703 & -13.799831 & 0.04 \\
56165.426100 & 334.798679 & -13.799841 & 0.04 \\
56165.427368 & 334.798683 & -13.799834 & 0.04 \\
56165.428443 & 334.798684 & -13.799849 & 0.04 \\
56166.447938 & 334.788135 & -13.803229 & 0.04 \\
56166.449013 & 334.788127 & -13.803240 & 0.04 \\
56166.509245 & 334.787477 & -13.803433 & 0.04 \\
56166.549260 & 334.787059 & -13.803579 & 0.04 \\
56167.385009 & 334.778460 & -13.806341 & 0.04 \\
56167.386086 & 334.778441 & -13.806314 & 0.04 \\	\enddata
\tablenotetext{a}{J2000 Coordinates}
\end{deluxetable}

\begin{landscape}
\begin{deluxetable}{llllllll} 
	\tabletypesize{\small}
	\tablecaption{Refined Orbital Elements\tablenotemark{a}\label{tab:elements}}
	\tablehead{
	\colhead{Object}  & \colhead{$a$~(AU)} & \colhead{$e$} & \colhead{$i$~(deg)} & \colhead{$\Omega$~(deg)} & \colhead{$\omega$~(deg)} & \colhead{$M$~(deg)} 
	}
	\startdata
(50000) Quaoar &$43.1881_{-22}^{+7}$&$0.037213_{-13}^{+6}$&$7.9948_{-1}^{+1}$&$189.004_{-4}^{+3}$&$162.72_{-2}^{+6}$&$276.06_{-6}^{+2}$&2MASS\\
&$0.0002$&$0.000004$&$0.0002$&$0.004$&$-0.01$&$0.01$&\\
(84522) 2002 TC302 &$55.763_{-11}^{+7}$&$0.2968_{-1}^{+7}$&$34.98787_{-7}^{+61}$&$23.8299_{-1}^{+2}$&$85.69_{-3}^{+14}$&$320.439_{-46}^{+2}$&2MASS\\
&$0.022$&$-0.0008$&$0.00058$&$0.0007$&$-0.22$&$0.081$&\\
(90377) Sedna&$543.4_{-2}^{+16}$&$0.85965_{-8}^{+46}$&$11.928240_{-18}^{+5}$&$144.428_{-3}^{+1}$&$310.944_{-8}^{+34}$&$358.215_{-1}^{+7}$&SDSS\\
&$-0.3$&$-0.00007$&$0.000057$&$0.007$&$-0.004$&$-0.001$&\\
(136199) Eris&$68.011_{-3}^{+2}$&$0.43578_{-2}^{+3}$&$43.8451_{-16}^{+8}$&$36.0509_{-4}^{+8}$&$150.80_{-1}^{+1}$&$202.76_{-4}^{+3}$&SDSS\\
&$0.000$&$-0.00000$&$0.0007$&$-0.0004$&$-0.00$&$0.01$&\\
(136472) Makemake&$45.488_{-1}^{+1}$&$0.16143_{-2}^{+1}$&$29.01313_{-8}^{+1}$&$79.2848_{-2}^{+12}$&$296.80_{-3}^{+2}$&$154.22_{-3}^{+4}$&SDSS\\
&$-0.003$&$0.00001$&$0.00012$&$-0.0020$&$-0.03$&$0.05$&\\
(202421) 2005 UQ513&$43.512_{-1}^{+1}$&$0.14391_{-6}^{+8}$&$25.72627_{-14}^{+4}$&$307.8351_{-9}^{+3}$&$220.02_{-6}^{+4}$&$221.72_{-6}^{+8}$&SDSS\\
&$-0.005$&$0.00001$&$-0.00022$&$-0.0012$&$0.04$&$-0.05$&\\
(225088) 2007 OR10&$66.96_{-1}^{+1}$&$0.50195_{-14}^{+9}$&$30.8137_{-8}^{+8}$&$336.8390_{-1}^{+1}$&$206.78_{-2}^{+1}$&$102.61_{-4}^{+6}$&2MASS\\
&$0.00$&$0.00000$&$0.0025$&$-0.0006$&$0.01$&$-0.02$&\\
	\enddata
	\tablenotetext{a}{Epoch of coordinates 56200.0 MJD. Offsets are with respect to the astDys nominal orbits on Sept. 1. 2012.}
	\tablenotetext{b}{Displayed uncertainties are in the last decimal place. Where necessary, multiple significant digits were included to reflect asymmetric uncertainties.}
	\tablenotetext{c}{Second row for each target displays the adjustment from the initial astDys elements.}
\end{deluxetable}
\end{landscape}

\begin{landscape}
\begin{deluxetable}{lllllll} 
	\tabletypesize{\small}
	\tablecaption{Notable Predictions \label{tab:predictions}}
	\tablehead{
	\colhead{ }  && \multicolumn{3}{c}{Star}&&\\
	\colhead{Object}  & \colhead{Date (UT)} & \colhead{R.A.\tablenotemark{a}} & \colhead{Dec.\tablenotemark{a}} & \colhead{Magnitude (r')} & \colhead{Shadow Velocity (km s$^{-1}$)} & \colhead{Vertical Shadow Uncertainty (km)}
	}
	\startdata
	2005 UQ513 & 2013/09/15 20:10 & 00:30:11.9 & 30:37:23.8 & 14.0 & 23.6 & 1000\\
	2005 UQ513 & 2014/11/19 02:53 & 00:29:40.5 & 30:41:23.3 & 15.9 & 21.9 & 2500\\
	Quaoar & 2013/07/01 21:42 & 17:35:13 & -15:23:33.9 & 17.5 & 24.5 & 1900\\
	Quaoar & 2013/07/09 02:41 & 17:34:40.5 & -15:23:37.5 & 14.4 & 23.3 & 1800\\
	Quaoar & 2013/07/12 20:54 & 17:34:24.2 & -15:23:43.2 & 12.9 & 22.8 & 1900\\
	Quaoar & 2014/09/10 02:35 & 17:37:32.3 & -15:31:08.9 & 18.0 & 4.2 & 3200\\
	Makemake & 2014/03/20 20:28 & 12:48:28.3 & 26:40:36.1 & 19.8 & 26.3 & 1430\\
	2007 OR10 & 2013/08/04 06:19 & 22:20:57.9 & -13:27:18.4 & 19.5 & 26.1 & 4300\\
	\enddata
	\tablenotetext{a}{J2000 Coordinates}
\end{deluxetable}
\end{landscape}

\begin{deluxetable}{lllll} 
	\tabletypesize{\small}
	\tablecaption{Observed and Predicted Offsets for (50000) Quaoar \tablenotemark{a}\label{tab:offsets}}
	\tablehead{
	\colhead{ }  & \multicolumn{2}{c}{RA (")} & \multicolumn{2}{c}{Dec (")}\\
	\colhead{Time (UT)}  & \colhead{Observed} & \colhead{Predicted} & \colhead{Observed} & \colhead{Predicted}
	}
	\startdata
	\multicolumn{5}{c}{2012-July-10}\\ \hline
	6.856185 & $4.04\pm0.1$ & 4.03 & $-0.00\pm0.1$ & 0.067\\
	6.895627 & $3.97\pm0.2$ & 3.94 & $-0.00\pm0.2$ & 0.066\\
	11.145036 & $-7.43\pm0.1$ & -7.63 & $-0.21\pm0.1$ & -0.12\\
	11.178836 & $-7.54\pm0.08$ & -7.72 & $-0.20\pm0.08$ & -0.12\\ \hline \\
	\multicolumn{5}{c}{2012-July-13}\\ \hline
	2.707929 & $4.49\pm0.06$ & 4.45 & $0.14\pm0.06$ & 0.14\\
	2.759136 & $4.36\pm0.08$ & 4.31 & $0.13\pm0.08$ & 0.13 \\
	6.324234 & $-4.88\pm0.2$ & -5.07 & $-0.12\pm0.2$ & -0.09\\
	6.375295 & $-5.02\pm0.15$ & -5.20 & $-0.11\pm0.15$ & -0.10\\
	\enddata
	\tablenotetext{a}{Offsets are determined from the element adjustment approach and are quoted as Quaoar-star.}
\end{deluxetable}

\end{document}